\documentclass[
    ,final            
  ]
  {aipproc}

\layoutstyle{6x9}

\usepackage{amssymb}

\begin{document}

\title{Charmonium and Charmonium--like (?) States}


\author{Kamal K. Seth}{
  address={Department of Physics and Astronomy, 
        Northwestern University, Evanston, IL, 60208, USA}
}

\begin{abstract}
The last few years have witnessed a renaissance in the spectroscopy of heavy quarks.  Several long elusive states have now been firmly identified, and several unexpected states have been reported by the high luminosity experiments at Belle, Babar, CLEO, and Fermilab.  These discoveries have posed important theoretical questions for our understanding of QCD, and a variety of theoretical models have been proposed.  These developments are critically discussed.
\end{abstract}

\maketitle


\section{Introduction}

Until a few years ago, in my talks I used to point out how heavy--quark ($c$, $b$) spectroscopy is so much cleaner experimentally than the spectroscopy of light quarks ($u$, $d$, $s$) because of narrow and well-separated states, and so much more amenable to understanding in terms of QCD because of the smaller value of the strong coupling constant, $\alpha_S$, and the less drastic relativistic effects.  As you will see, this was only true as long as we dealt only with bound states.   It is not true now, as we have moved on to higher states.

\section{Charmonium $(c\bar{c})$}

The spectrum of charmonium states is well known.  Below the $D\overline{D}$ threshold at 3730 MeV, the bound states are $1^1S_0(\eta_c)$, $1^3S_1(J/\psi)$, $2^1S_0(\eta_c')$, $2^3S_1(\psi')$, $1^1P_1(h_c)$, and $1^3P_J(\chi_{c0,c1,c2})$.   Despite thirty years of spectroscopy, two glaring holes have remained in the spectrum of the bound states of charmonium.  Neither SLAC, nor Fermilab, nor BES were able to identify the two spin--singlet states, the $\eta_c'(2^1S_0)$ and $h_c(1^1P_1)$, both known to be bound states.  A milestone in charmonium spectroscopy has now been reached.  Both these states have now been firmly identified.

\subsection{$\eta_c'(2^1S_0)$ -- The Radial Excitation of the Charmonium Ground State}

In 1982, the Crystal Ball Collaboration at SLAC claimed the identification of $\eta_c'$ in radiative transition from $\psi'$, with mass $M(\eta_c')=3594(5)~\mathrm{MeV}$.  This corresponded to the $2S$ hyperfine splitting $\Delta M_{hf}(2S)\equiv M(\psi')-M(\eta_c')=92(5)~\mathrm{MeV}$.  This was rather surprising because a `model--independent' prediction based on $\Delta M_{hf}(1S)=117(1)~\mathrm{MeV}$ \cite{pdg}, is that $\Delta M_{hf}(2S)=62(2)~\mathrm{MeV}$.  Fortunately, the Crystal Ball observation was never confirmed; $\eta_c'$ remained unidentified despite repeated attempts by the $\bar{p}p$ experiment E760/E835 at Fermilab, and the $e^+e^-$ measurements by DELPHI, L3, and CLEO.

The first observation of $\eta_c'$ was reported by Belle in $B$-decays \cite{belle-etacp}.  This was followed by identifications by CLEO \cite{cleo-etacp} and BaBar \cite{babar-etacp} in two--photon fusion, $\gamma\gamma\to\eta_c'\to K_SK\pi$, as illustrated in Fig.~1 (left).  Since then, both Belle and BaBar have reported its observation in double charmonium production in $e^+e^-$ collisions  in the $\Upsilon(4S)$ region.  The weighted average of the observed masses is $M(\eta_c')=3638(4)~\mathrm{MeV}$ \cite{pdg}, which leads to the $2S$ hyperfine splitting, $\Delta M_{hf}(2S)=48(4)~\mathrm{MeV}$. This is unexpectedly small compared to $\Delta M_{hf}(1S)=117(1)~\mathrm{MeV}$.  Attempts have been made to explain this by invoking channel mixing, but it is fair to say that the observation remains a challenge to the theorists.  The width of $\eta_c'$ remains essentially undetermined, so far. It is hoped that with the $\sim30$ million $\psi'$ that CLEO-c expects to have soon, the direct M1 transition, $\psi'\to\gamma\eta_c'$ can be identified, and the width and mass of $\eta_c'$ can be determined with precision.

\begin{figure}[!tb]
\includegraphics[width=3.0in]{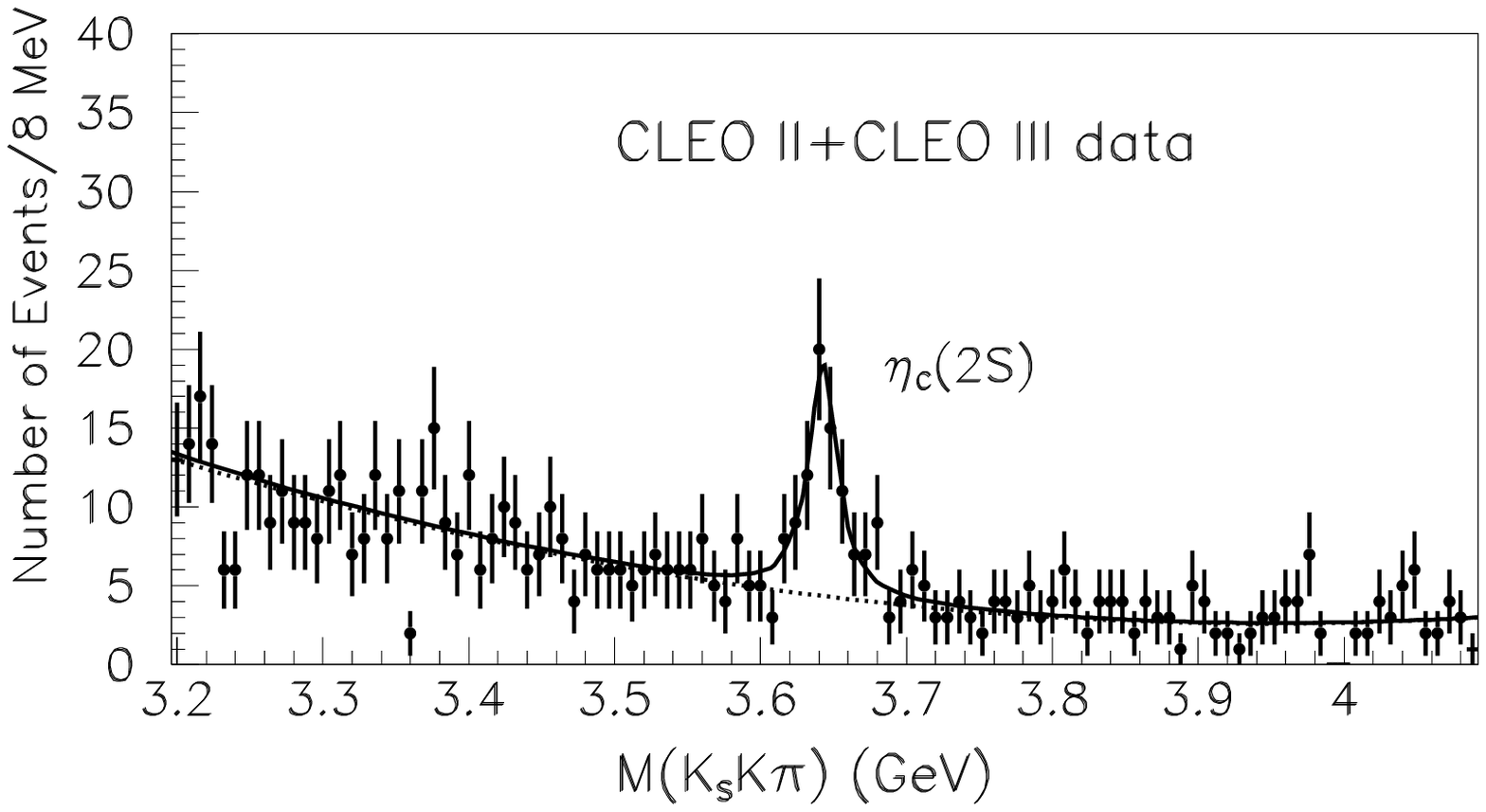}
\raisebox{1.8in}{\rotatebox{270}{\includegraphics[width=1.7in]{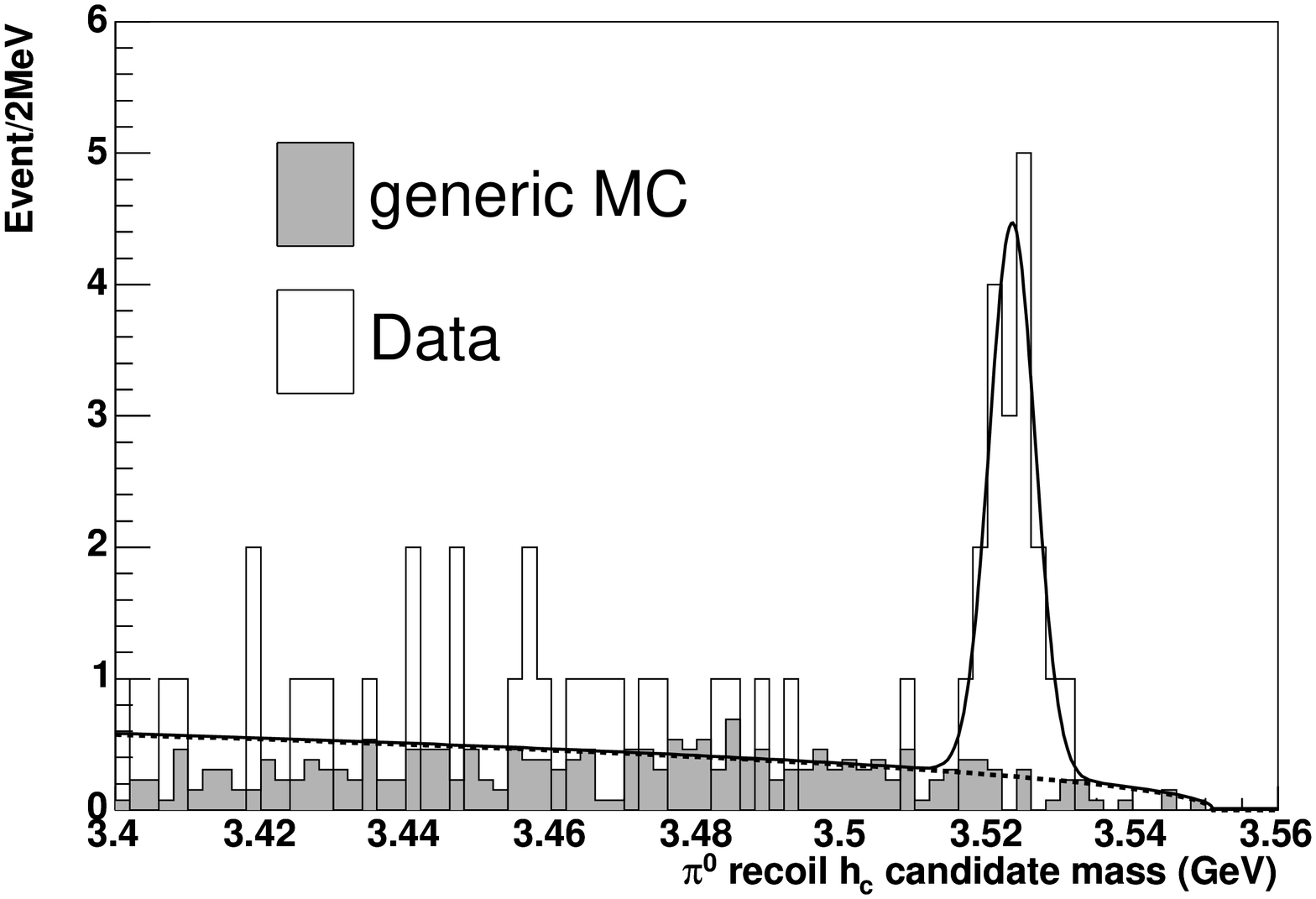}}}
\caption{(left) Observation of $\eta_c'$ in the reaction $\gamma\gamma\to K_SK\pi$ by CLEO \cite{cleo-etacp}.  (right)  Observation of $h_c$ in exclusive analysis of $\psi'\to \pi^0h_c,~h_c\to\gamma\eta_c$ by CLEO \cite{cleo-hc}.}
\end{figure}

\subsection{$h_c(1^1P_1)$ -- The Singlet $P$--state of Charmonium}

This state has been the object of many frustrating searches since the early days of charmonium spectroscopy.  The great interest in identifying it comes from the fact that if the $|q\bar{q}>$ confinement potential is a Lorenz scalar, as is generally assumed, there is no long-range spin-spin interaction, and for $P$--wave (and higher $L$--waves), the hyperfine splitting should be identically zero, i.e., $\Delta M_{hf}(1P)\equiv M(\left<^3P_J\right>) - M(^1P_1)=0$.  The weighted average of the masses of the $^3P_J$ states, $\chi_{c0},~\chi_{c1},~\chi_{c2}$, is very accurately known, being
$$M(\left<^3P_J\right>) = [5M(\chi_{c2})-3M(\chi_{c1})-M(\chi_{c0})]=3525.36(6)~\mathrm{MeV}$$
so that $M(^1P_1)$ should be exactly the same.  However, speculations abound on how it could be up to 10--15 MeV different from this.  A precision measurement of $M(h_c)$ is therefore mandatory.

The main difficulty in identifying $h_c$ is that its formation in radiative decay of $\psi'$ is forbidden by charge conjugation, as is its radiative decay to $J/\psi$.  Attempts by the Fermilab E760/E835 to search for $h_c$ via the reaction $p\bar{p}\to h_c\to \pi^0 J/\psi$ were unsuccessful, as were earlier attempts by the Crystal Barrel Collaboration to search for it.

With a state-of-the-art detector and large luminosity, CLEO has returned to the isospin forbidden reaction $\psi'\to \pi^0h_c,~h_c\to\gamma\eta_c$, and successfully identified $h_c$ \cite{cleo-hc}.  In the inclusive measurements, either the photon energy or the $M(\eta_c)$ were loosely constrained, and $h_c$ was identified as an enhancement in the $\pi^0$ recoil spectrum.  In the exclusive measurement, shown in Fig.~1 (right), neither the photon energy nor $M(\eta_c)$ were constrained, but $\eta_c$ was identified in seven hadronic decays. The two measurements gave consistent results, their average being $M(h_c)=3524.4\pm0.6\pm0.4~\mathrm{MeV}$, which leads to $\Delta M_{hf}(1P)=+1.0\pm0.6\pm0.4~\mathrm{MeV}$.  This time the surprise is a pleasant one, in that the naive expectation of zero hyperfine splitting seems to be almost true.  Once again, it is hoped that with the nearly ten times $\psi'$ which are expected to be soon available at CLEO, a better measurement of $M(h_c)$ and $\Gamma(h_c)$ will be forthcoming soon.

\section{The Surprising and Unexpected Charmonium--like (?) States}

During the last two years, unexpected states have been popping up all over.  The first of these is X(3872), and the last is Y(4260).  In between are three states X, Y, Z, all having masses near 3940 MeV.  This proliferation is both exciting and rather baffling.  It arises primarily from the fact that with huge integrated $e^+e^-$ luminosities available at the $B$--factories, the Belle and BaBar detectors are observing very weakly excited resonances.  It is obvious that it will take a while before the dust settles down, and when it does, it is likely that not all the resonances will survive.

Of the alphabet soup, there are only two resonances which have been observed by more than one experiment.  These are X(3872) and Y(4260).  The other three, X, Y, Z(3940) have been reported only by Belle, and the silence from BaBar is deafening.

\textbf{X(3872):}  First reported by Belle \cite{belle-x}, this resonance, which decays primarily into $\pi^+\pi^-J/\psi$, has been confirmed by CDF, D\O, and BaBar.  Its average mass is $M(\mathrm{X})=3871.5\pm0.4~\mathrm{MeV}$, and width $\left<\Gamma\right>\le2.3~\mathrm{MeV}$.  A variety of theoretical explanations for X(3872) have been suggested, ranging from mixed charmonium to a $D\overline{D}^*$ molecule.  A large number of decays of X(3872) have been investigated and angular correlations have been studied.  My summary of these is that its $J^{PC}=1^{++}$ or $2^{-+}$.  It could be a displaced $1^1D_2(2^{-+})$ or $2^3P_1(1^{++})$ state of charmonium, or a $D\overline{D}^*$ molecule.  If it is the latter, the binding energy of the molecule is $E_b=+0.61\pm0.62~\mathrm{MeV}$ \cite{cleo-dmass}.  This produces a big problem for the molecule, because Swanson \cite{swanson} predicts the ratio $R\equiv\Gamma(\mathrm{X}\to D\overline{D}\pi^0)/\Gamma(\mathrm{X}\to \pi^+\pi^-J/\psi)\approx1/20$, while Belle \cite{belle-xd} has measured $R\approx10$.

\textbf{Y(4260), or V(4260):}  BaBar \cite{babar-y4260} has reported an enhancement in the $\pi^+\pi^-J/\psi$ invariant mass, labeled Y(4260), in $e^+e^-$ annihilation following initial state raditiation (ISR).  They report $M(\mathrm{Y})=4259\pm8^{+2}_{-6}~\mathrm{MeV}$ and $\Gamma(\mathrm{Y})=88\pm23^{+6}_{-4}~\mathrm{MeV}$.  The production of this state in ISR would make it a vector (hence my suggestion to call it V(4260)).  Unfortunately, all the charmonium vectors in the 3.8--4.4 GeV mass region are spoken for, and actually there is a deep minimum in $R\equiv\sigma(e^+e^-\to\mathrm{hadrons})/\sigma(e^+e^-\to\mu^+\mu^-)$ \cite{seth}.  thus, Y(4260) is extremely surprising and its existence had to be independently confirmed.  CLEO \cite{cleo-y4260} has now done that. Although CLEO statistics are much smaller, the much lower background enables it to make a firm confirmation of the Y(4260) resonance with  $M(\mathrm{Y})=4284^{+17}_{-16}\pm4~\mathrm{MeV}$ and $\Gamma(\mathrm{Y})=73^{+39}_{-25}\pm5~\mathrm{MeV}$.  Fig.~2 shows the spectra obtained by both BaBar and CLEO.  No theoretical understanding of Y(4260) exists so far.  It is a truely mysterious state.

\begin{figure}[!tb]
\includegraphics[width=2.2in]{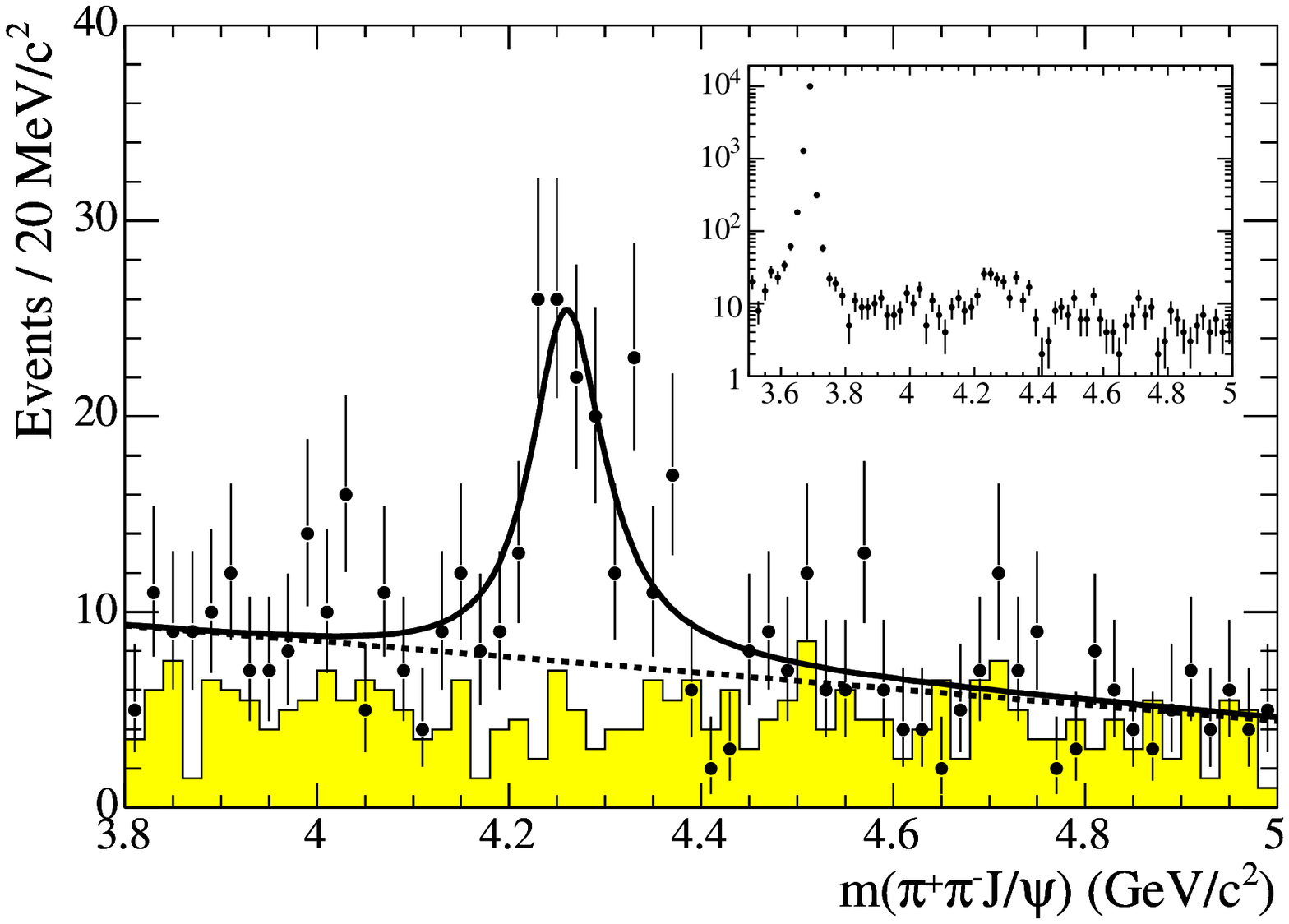}
\includegraphics[width=2.2in]{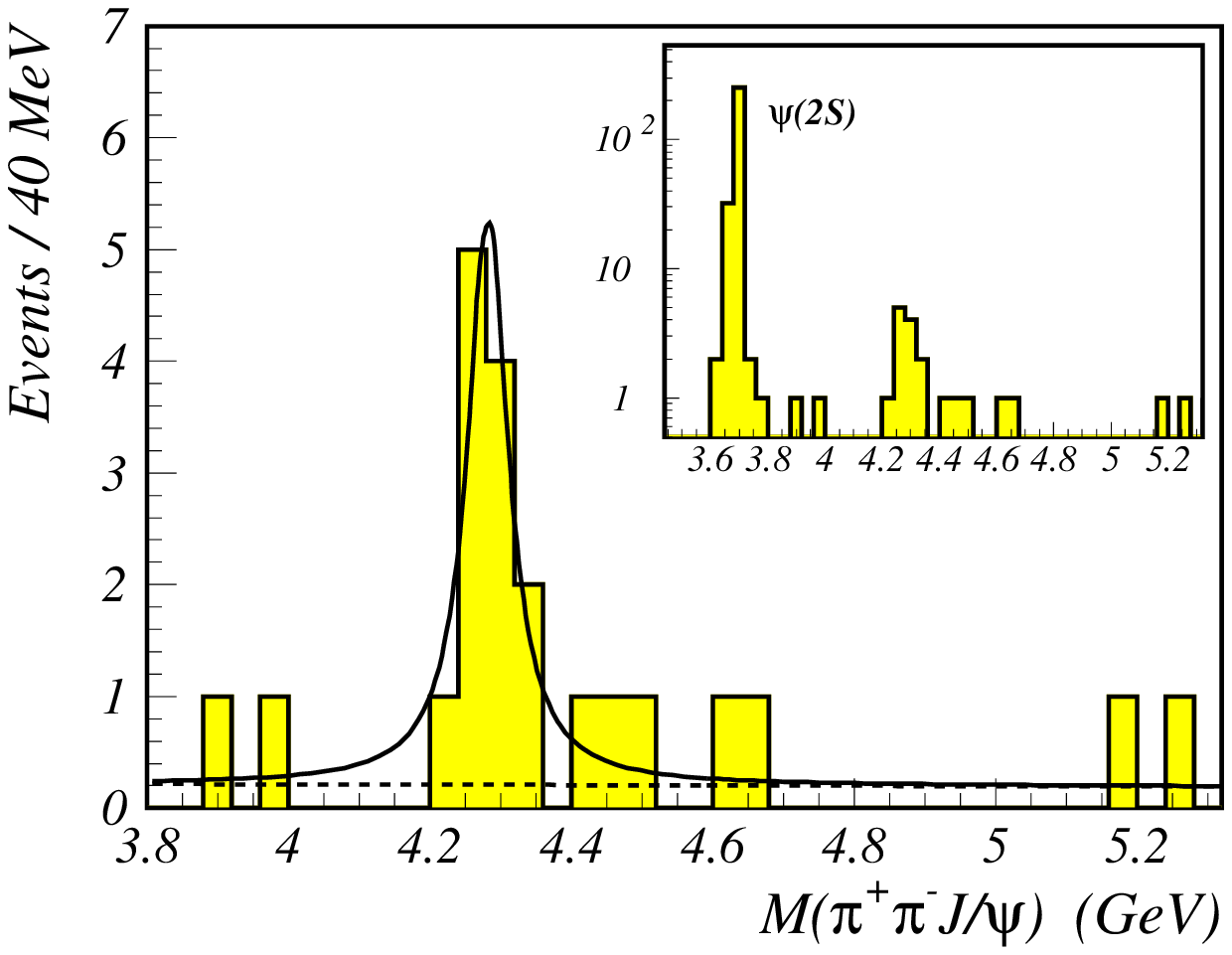}
\caption{Observation of Y(4260) by BaBar (left) and CLEO (right) in the spectra of $M(\pi^+\pi^-J/\psi)$.}
\end{figure}

\begin{figure}[!tb]
\includegraphics[width=2.in]{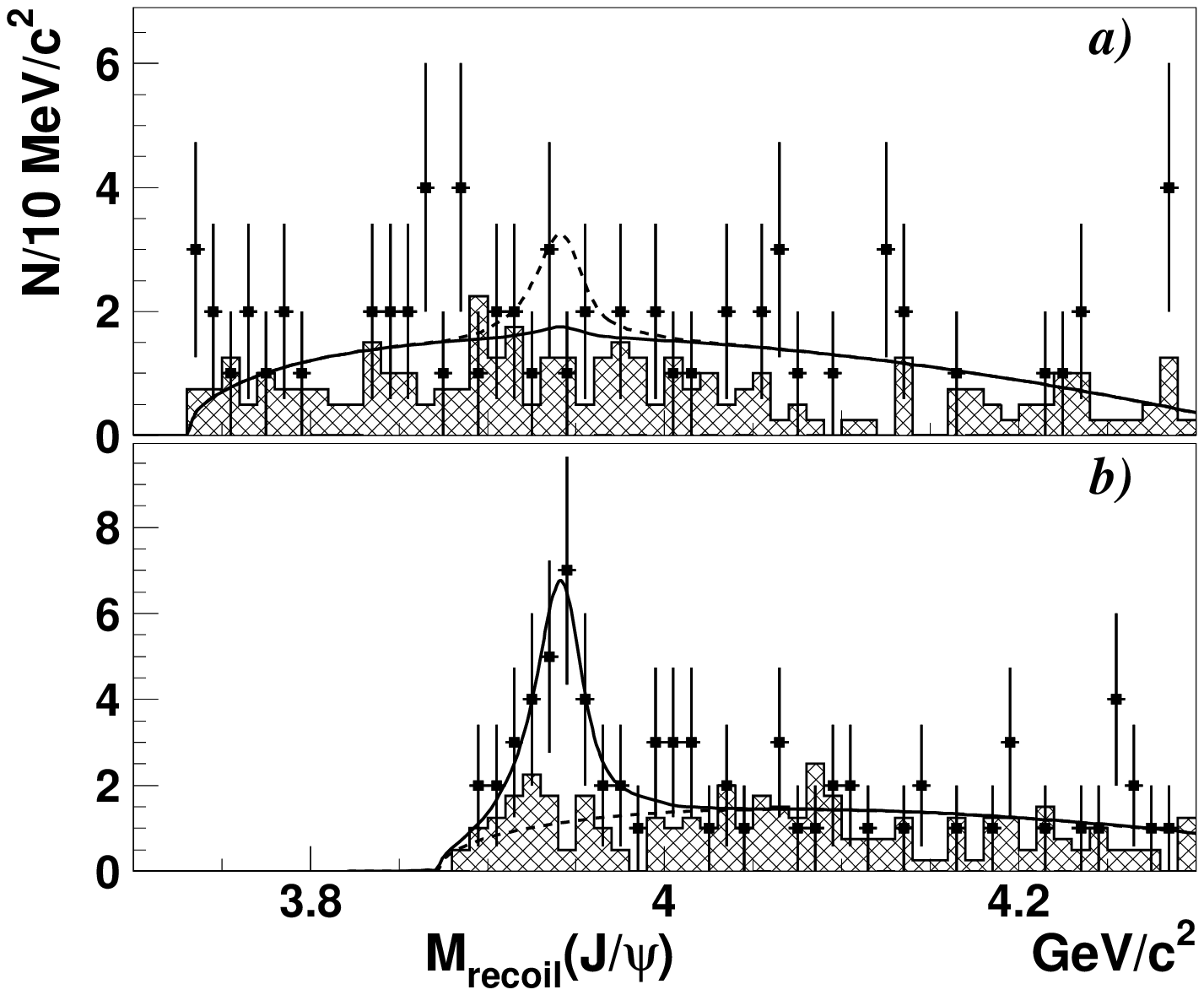}
\includegraphics[width=1.6in]{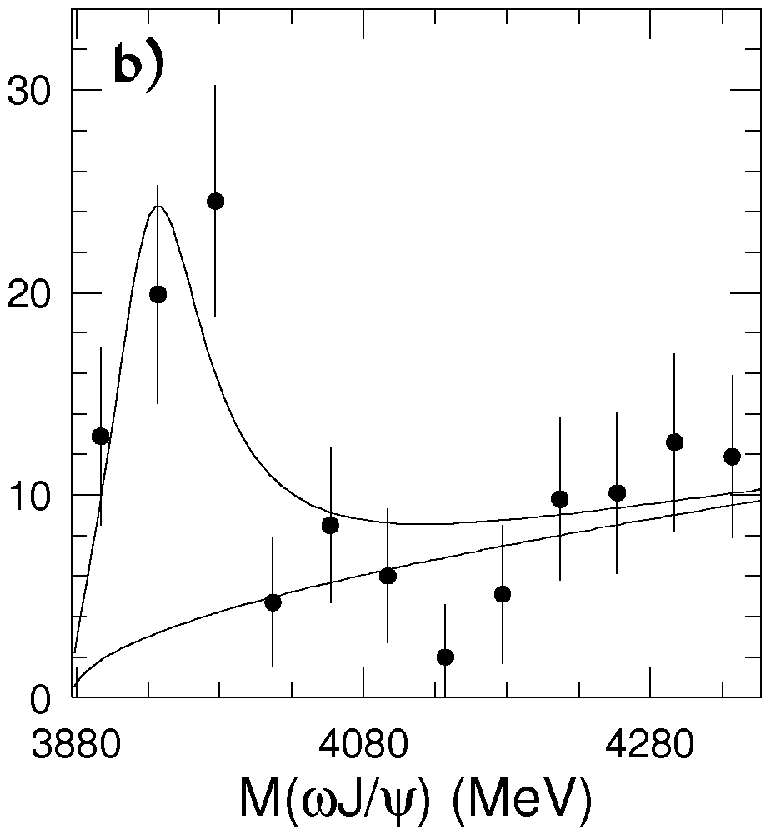}
\includegraphics[width=1.6in]{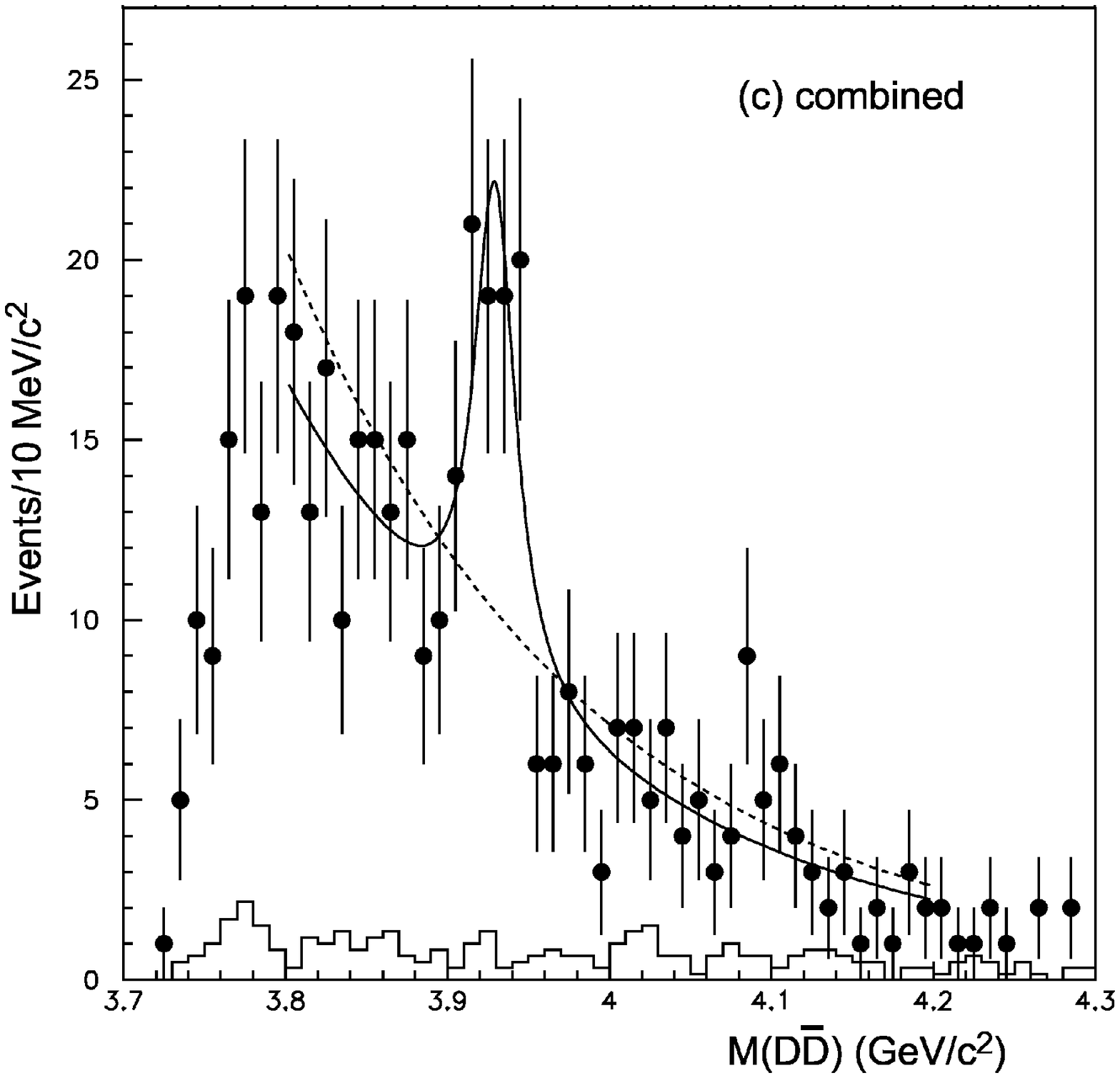}
\caption{The three resonances of Belle: (left) X(3936), (middle) Y(3943), and (right) Z(3929).}
\end{figure}

\begin{table}[!tb]

\begin{footnotesize}
\begin{tabular}{|c|c|c|c|c|c|c|c|}
\hline
 &  M  & $\mathcal{L}$  & N (evts) & $\Gamma$  & Formed in/ & No decays to & Suggested? \\
 & (MeV) & fb$^{-1}$ & & (MeV)  & Decays to & & \\
\hline
X \cite{belle-x3940} & 3936(14) & 357 & 266(63) & 39(26) & $e^+e^-\to J/\psi(\mathrm{X})$ & $\mathrm{X}\nrightarrow \overline{D}D$ & $\eta_c''(2^1S_0)$ \\
 & & & 25(7) & 15(10) & $\mathrm{X}\to\overline{D}D^*$ & $\mathrm{X}\nrightarrow\omega J/\psi$  & \\
Y \cite{belle-y3940} & 3943(17) & 253 & 58(11) & 87(34) & $B\to K\mathrm{Y}$ & & $c\bar{c}$ hybrid \\
 & & & &  & $\mathrm{Y}\to \omega J/\psi$ & $\mathrm{Y}\nrightarrow \overline{D}D^*$ &  \\
Z \cite{belle-z3940} & 3929(5) & 395 & 64(18) & 29(10) & $\gamma\gamma\to\mathrm{Z}$ &  & $\chi_{c2}'(2^3P_2)$\\
 &  & & & & $\mathrm{Z}\to D\overline{D}$  & $\mathrm{Z}\nrightarrow \overline{D}D^*$ & \\
\hline 
\end{tabular}
\end{footnotesize}
\caption{Summary of the Belle resonances X, Y, Z.}
\end{table}

\textbf{X, Y, Z(3940):} Belle \cite{belle-x3940,belle-y3940,belle-z3940} has reported three different states, all having the same mass within $\pm6~\mathrm{MeV}$, but produced in different reactions and decaying into different final states (see Fig.~3).  The observations are summarized in Table~1.  The statistics of the observations are low, and there are open questions.  Since Estia Eichten \cite{eichtentalk} is talking about these states in a plenary talk, I will not go into details here.  I do, however, want to note some of the problems with the favorite theoretical suggestions (listed in the last column of Table 1) for the possible nature of these states.

\begin{itemize}
\item \textbf{X(3940):} Can this be $\eta_c''(3^1S_0)$?  $\psi(4040)$ is generally accepted as $\psi''(3^3S_1)$.  If X(3940) is $\eta_c''$, $\Delta M_{hf}(3S)\approx100~\mathrm{MeV}$.  With $\Delta M_{hf}(2S)=48(4)~\mathrm{MeV}$, can we have $\Delta M_{hf}(3S)\approx100~\mathrm{MeV}$?  Further, is it possible that X(3940) is the same as Z(3929)?  $D\overline{D}$ and $D\overline{D}^*$ decays of both need to be investigated carefully.

\item \textbf{Y(3940):}  Can this be a hybrid?  The lowest $|c\bar{c}g>$ hybrid is predicted with mass $M=4300-4500~\mathrm{MeV}$.  $D\overline{D}$ and $D\overline{D}^*$ decays need to be measured.

\item \textbf{Z(3929):}  Why is there yield only in $D^0\overline{D}^0$, and none in $D^+D^-$?  Why no $D\overline{D}^*$?
\end{itemize}

My biased conclusion is that it is entirely likely that not all three X, Y, Z are seperate entities.  We need consistent analysis of the full Belle data set, and for BaBar to weigh in with confirmations or refutations.






\bibliographystyle{aipproc}   


\end{document}